\documentclass[11pt]{article}
\usepackage{axodraw}
\usepackage{graphicx}
\usepackage{enumerate}
\def\lsim{\mathrel{\vcenter{\hbox{$<$}\nointerlineskip\hbox{$\sim$}}}}
\def\gsim{\mathrel{\vcenter{\hbox{$>$}\nointerlineskip\hbox{$\sim$}}}}
\def\egzk{E_{\rm GZK}}

\def\nue{\nu_e}

\def\nuebar{\bar{\nu}_e}

\newcommand{\be}{\begin{equation}}
\newcommand{\ee}{\end{equation}}
\newcommand{\ba}{\begin{eqnarray}}
\newcommand{\ea}{\end{eqnarray}}

\def\lsim{\raise0.3ex\hbox{$\;<$\kern-0.75em\raise-1.1ex\hbox{$\sim\;$}}}
\def\gsim{\raise0.3ex\hbox{$\;>$\kern-0.75em\raise-1.1ex\hbox{$\sim\;$}}} 

\newcommand{\mx}{\left[\begin{array}}
\newcommand{\finmx}{\end{array}\right]} 
\newcommand{\mxp}{\left(\begin{array}} 
\newcommand{\finmxp}{\end{array}\right)} 
\def\beq{\begin{equation}}
\def\eeq{\end{equation}}
\def\bea{\begin{eqnarray}}
\def\eea{\end{eqnarray}}

\def\mathbf#1{\hbox{\bf #1}}
\def\textrm#1{\hbox{#1}}

\def\lsim{\raise0.3ex\hbox{$\;<$\kern-0.75em\raise-1.1ex\hbox{$\sim\;$}}}
\def\gsim{\raise0.3ex\hbox{$\;>$\kern-0.75em\raise-1.1ex\hbox{$\sim\;$}}}
\newcommand {\ignore}[1]{}

\begin{document}
\vspace*{-1in}
\renewcommand{\thefootnote}{\fnsymbol{footnote}}
\begin{flushright}
\texttt{
} 
\end{flushright}
\vskip 5pt
\begin{center}
{\Large{\bf Absolute neutrino mass update}\footnote{Talk presented 
by H.~P\"as at SUSY02, 10th International Conference on 
\it{Supersymmetry and Unification of Fundamental Interactions}
\rm, 17-23/06/02, DESY, Hamburg
}}
\vskip 25pt

{\sf 
Heinrich P\"as$^a$,  
Thomas J. Weiler$^b$
}
\vskip 10pt
{\it \small $^a$ Institut f\"ur Theoretische Physik und Astrophysik\\
Universit\"at W\"urzburg\\ D-97074 W\"urzburg, Germany}\\

\vskip 10pt
{\it \small $^b$ Department of Physics and Astronomy, Vanderbilt University,\\
Nashville, TN 37235, USA}  

\vskip 20pt

{\bf Abstract}
\end{center}

\begin{quotation}
{\small 
The determination of absolute neutrino masses is crucial for the 
understanding of theories underlying the standard model, such as SUSY.
We review the experimental prospects to determine
absolute neutrino masses and the correlations among approaches,
using the $\Delta m^2$'s inferred from neutrino oscillation experiments
and assuming a three neutrino Universe.
}
\end{quotation}

\vskip 20pt  

\setcounter{footnote}{0}
\renewcommand{\thefootnote}{\arabic{footnote}}


\section{Neutrinos and new physics}

The most pending puzzles in particle and astroparticle physics concern
the origin of mass, the unification of interactions, the nature of the
dark matter in the universe, the existence of hidden extra dimensions, 
the origin of the highest energy cosmic rays and the explanation of the 
matter-antimatter excess. 
The investigation of the unknown absolute neutrino mass scale 
is situated at a 
crossing point of these tasks: 
\begin{itemize}

\item
The most elegant explanation for
light neutrino masses is the see-saw mechanism, in which a large Majorana 
mass scale $M_R$ drives the light neutrino masses down to or below the 
sub-eV scale,
\be{}
m_{\nu}=m_D^2/M_R,
\ee
where the Dirac neutrino masses are typically of the order of the weak scale.
A combination of information about $m_D$ from charged lepton flavor violation
mediated by sleptons
(see e.g. \cite{dprr}) and $m_{\nu}$ may allow to probe the scale $M_R$ not 
far from the GUT scale.

\item
An alternative mechanism generates neutrino masses radiatively 
at the SUSY scale,
with 
R-parity violating couplings $\lambda^{(')}$,
fermions $f$ and squarks or sleptons in the loop,
\be{}
m_{\nu} \propto \lambda^{(')} \lambda^{(')} m_f^2/ (16 \pi^2 M_{SUSY}).
\ee
In this case information about the strength of couplings and the masses of 
SUSY partners can be obtained from
absolute neutrino masses (see e.g. \cite{bkp}).

\item
In theories with large extra dimensions small neutrino masses may be generated
by volume-suppressed  couplings to right-handed neutrinos which can
propagate in the bulk, by the breaking of lepton number on 
a distant brane, or by the curvature of the extra dimension.
Thus neutrino masses can provide information about the volume or 
the geometry of the large extra dimensions (see e.g. 
\cite{xtranus}). 

\item
A
simple and elegant explanation of the matter-antimatter excess in the 
universe is given by the out-of-equilibrium decay of heavy Majorana 
neutrinos in leptogenesis scenarios. 
To avoid strong washout processes of the generated
lepton number asymmetry light neutrino masses with $m_{\nu}<0.2$~eV are 
required \cite{buchm}. 

\end{itemize}

In fact it is a true experimental challenge
to determine an absolute neutrino mass below 1~eV.
Three approaches 
have the potential to accomplish the task, namely  
larger versions of the tritium end-point distortion measurements,
limits from the evaluation of the large scale structure in the universe,
and 
next-generation neutrinoless double beta decay ($0\nu\beta\beta$) experiments.
In addition there is a fourth possibility: 
the extreme-energy cosmic-ray experiments 
in the context of the recently emphasized Z-burst model.
For discussions of the sensitivity in time of flight measurements
of supernova (${\cal O}$(1~eV))
or gamma ray burst neutrinos (${\cal O}(10^{-3}$~eV), 
assuming complete understanding of
GRB's and large enough rates), see \cite{BBM00}.

\section{Tritium beta decay}

In tritium decay, the larger the mass states comprising $\nuebar$,
the smaller is the Q-value of the decay.
The manifestation of neutrino mass is a reduction of phase space
for the produced electron at the high energy end of its spectrum.
An expansion of the decay rate formula about $m_{\nue}$ leads to
the end point sensitive factor 
\be{}
m^2_{\nu_e}\equiv \sum_j\,|U_{ej}|^2\,m^2_j\,,
\ee
where the sum is over mass states $m_i$ which can kinematically alter
the end-point spectrum.
If the neutrino masses are nearly degenerate,
then unitarity of the mixing matrix $U$ leads immediately to a bound on
$\sqrt{m^2_{\nu_e}}=m_3$.
A larger tritium decay experiment (KATRIN) to reduce 
the present 2.2~eV $m_{\nu_e}$ bound is planned to start taking data in 2007;
direct mass limits as low as 0.4~eV, or even 0.2~eV, may be possible
in this type of experiment \cite{katrin}.

\section{Cosmological limits}

In the currently favored $\Lambda$DM cosmology,
there is scant room left for the neutrino component.
The power spectrum of early-Universe density perturbations
is processed by gravitational instabilities.
However, 
the free-streaming relativistic 
neutrinos suppress the growth of fluctuations
on scales below the horizon 
(approximately the Hubble size $c/H(z)$) 
until they become nonrelativistic at 
$z\sim m_j/3T_0 \sim 1000\,(m_j/{\rm eV})$ (for an overview see 
\cite{king}).

A recent limit \cite{elg} 
derived from the 2dF Galaxy Redshift Survey power spectrum
constrains the fractional contribution of massive neutrinos to the total
mass density to be less than 0.13, translating into a bound on the
sum of neutrino mass eigenvalues,  $\sum_j m_j<1.8$~eV (for a total
matter mass density $0.1<\Omega_m<0.5$ and a scalar spectral index $n=1$).
A limit from gravitational lensing by dwarf satellite galaxies 
reveals sufficient structure to limit
$\sum_j m_j<0.74$~eV, under some reasonable but unproven assumptions
\cite{dalal}.
In ref. \cite{hannestad} it has been shown, that a combination
of Planck satellite CMB data with the SDSS sky survey will improve the
sensitivity
down to $\sum_j m_j= 0.12$~eV. A future sky survey with an order of magnitude 
larger survey volume would allow to reach even $\sum_j m_j= 0.03-0.05$~eV.

Some caution is warranted in the cosmological approach to neutrino mass,
in that the many cosmological parameters may conspire in 
various combinations to yield nearly identical CMB and large scale structure 
data.
An assortment of very detailed data may be needed to resolve 
the possible ``cosmic ambiguities''.

\section{Neutrinoless double beta decay}

The $0\nu\beta\beta$ rate is a sensitive tool for the
measurement of the absolute mass-scale for Majorana neutrinos \cite{kps}.
The observable measured in the amplitude of $0\nu\beta\beta$ 
is the $ee$ element of the neutrino mass-matrix in the flavor basis.
Expressed in terms of the mass eigenvalues and 
neutrino mixing-matrix elements, it is 
\be{}
m_{ee}= |\sum_i U_{ei}^2 m_i|\,.
\label{dbeqn}
\ee
A reach as low as $m_{ee}\sim 0.01$~eV seems possible 
with double beta decay projects under preparation such as 
GENIUSI, MAJORANA, EXO, XMASS or MOON. 
This provides a substantial improvement over the current bound from the IGEX 
experiment,
$m_{ee}< 0.4$~eV \cite{igex}. A recent evidence claim \cite{evi} 
by the Heidelberg-Moscow experiment
reports a best fit value of $m_{ee}=0.4$~eV,
but is subject to possible systematic uncertainties.

For masses in the interesting range $\gsim 0.01$~eV, 
the two light mass eigenstates are nearly degenerate and so the 
approximation $m_1 =m_2$ is justified.
Due to the restrictive CHOOZ bound, $|U_{e3}|^2 < 0.025$,
the contribution of the third mass eigenstate 
is nearly decoupled from $m_{ee}$ and so
$U^2_{e3}\,m_3$ may be neglected in the $0\nu\beta\beta$ formula.
We label by $\phi_{12}$ the relative phase between
$U^2_{e1}\,m_1$ and $U^2_{e2}\,m_2$.
Then, employing the above approximations,
we arrive at a very simplified expression for $m_{ee}$:
\be{}
m^2_{ee}=\left[1-\sin^2 (2\theta_{\rm sun})\,
       \sin^2 \left(\frac{\phi_{12}}{2}\right)\right]\,m^2_1\,.
\label{dbeqn2}
%
\ee
The two CP-conserving values of $\phi_{12}$ are 0 and $\pi$.
These same two values give maximal constructive and destructive
interference of the two dominant terms in eq.\ (\ref{dbeqn}),
which leads to upper and lower bounds for the observable
$m_{ee}$ in terms of a fixed value of $m_1$, 
$\cos (2\theta_{\rm sun})\;m_1 \leq m_{ee} \leq m_1$
with 
$\cos (2\theta_{\rm sun}) \gsim 0.1$ weakly bounded  
for the LMA solution \cite{con}. This uncertainty disfavors $0\nu\beta\beta$
in comparison to direct measurements if a specific value of $m_1$
has to be 
determined, while $0\nu\beta\beta$ is more sensitive as long as 
bounds on $m_1$ are aimed at. 
Knowing the value of $\theta_{\rm sun}$ better will improve
the estimate of the inherent uncertainty in $m_1$.
For the LMA solar solution, the 
forthcoming Kamland experiment should reduce the error in the 
mixing angle $\sin^2 2 \theta_{\rm sun}$ to $\pm 0.1$ \cite{barger00}.

\begin{figure}
\centerline{\resizebox{9cm}{7cm}{\includegraphics{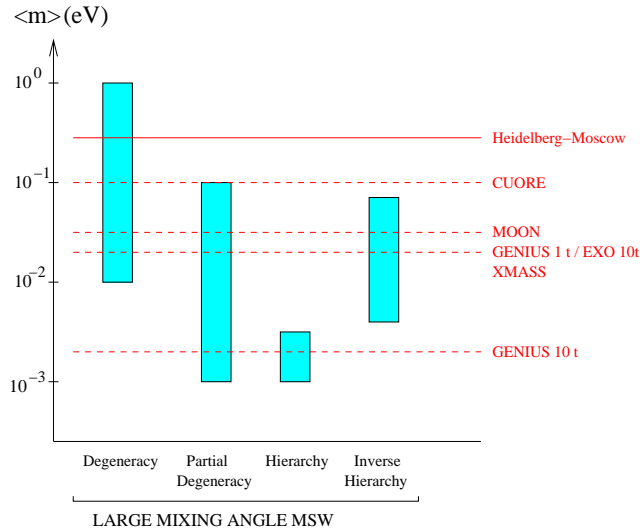}}}
 \caption{Different neutrino mass spectra versus sensitivities of future
double beta decay projects. A futuristic 10 ton Genius experiment may
test all neutrino spectra.
}
 \label{fig:cstates}
\end{figure}

In the far future,
another order of magnitude in reach 
is available to the 
10 ton version of GENIUS, should it be funded and commissioned.
Such a project would be sensitive to all different kinds of neutrino 
spectra including hierarchical ones, a summary is given in fig. 
\ref{fig:cstates}.

\begin{figure}[t]
\vspace*{-2.5cm} 
\epsfxsize=10cm %
\begin{center} 
\epsfbox{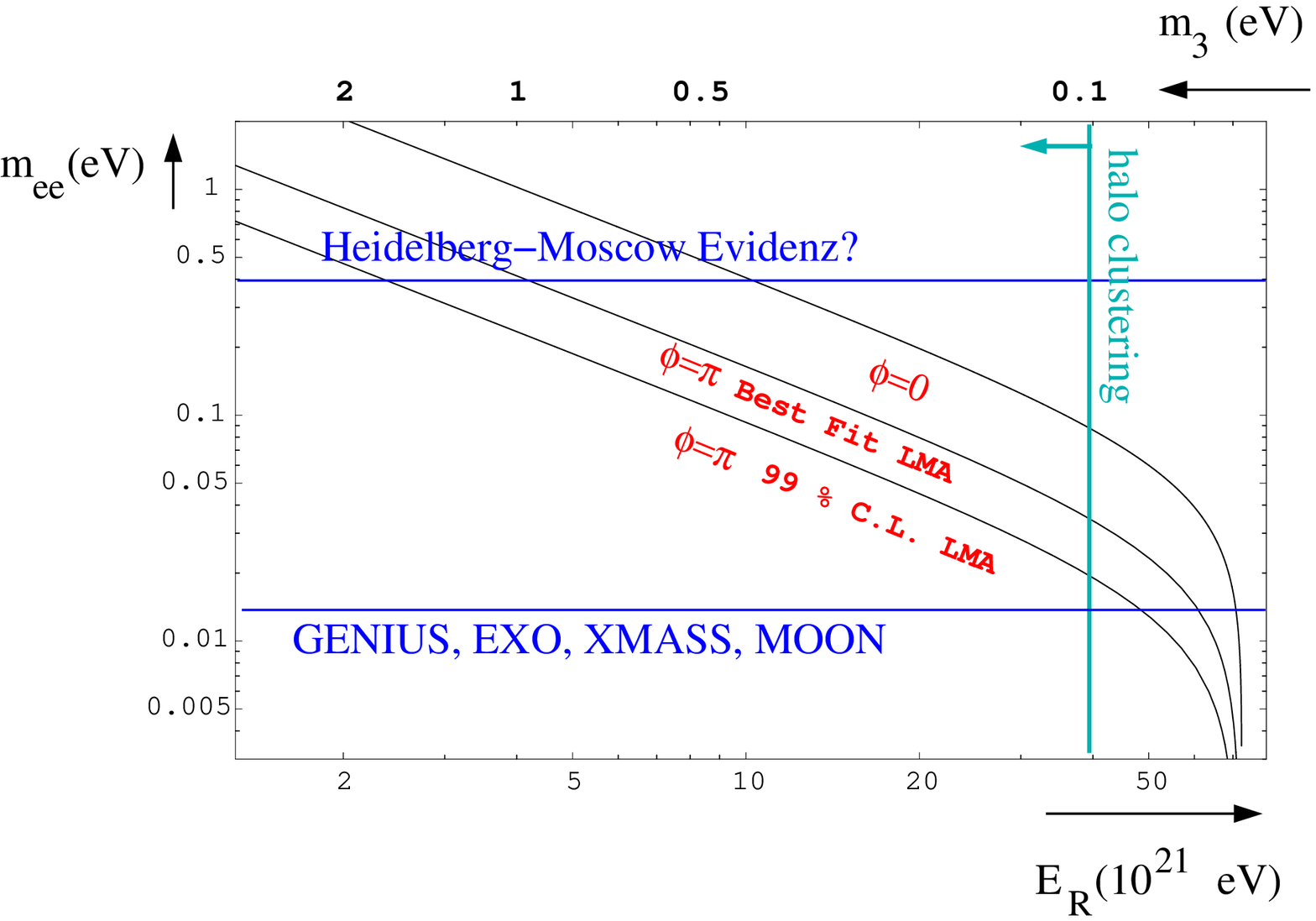}
\end{center} 
\vspace*{-3.7cm} 
\caption{$0\nu\beta\beta$ observable $m_{ee}$ 
versus mass of the heaviest neutrino
$m_3$, or, alternatively, the resonant Z-burst energy $E_R$.
The curved lines show allowed regions for different solutions of the solar 
neutrino anomaly; from top to bottom, the case for $\phi_{12}=0$,
$\phi_{12}=\pi$ for the best fit and the 99 \% C.L. range of the
LMA solution.
The region between the $\phi_{12}=0$ 
and the $\phi_{12}=\pi$ lines are allowed in the various solar 
solutions. The straight lines correspond to the Heidelberg--Moscow 
evidence and the sensitivity of next generation $0\nu\beta\beta$ projects.
  \label{uhe}}
\end{figure}

\section{Extreme energy cosmic rays in the Z-burst model}

It was expected that the extragalactic spectrum 
would reveal an end at the 
Greisen-Kuzmin-Zatsepin (GZK) cutoff energy of
$\egzk \sim 5\times 10^{19}$~eV. 
The origin of the GZK cutoff is the degradation of nucleon energy by the 
resonant scattering process $N+\gamma_{2.7K}\rightarrow \Delta^*
\rightarrow N+ \pi$ when the nucleon is above the resonant threshold $\egzk$.
The concomitant energy-loss factor is
$\sim (0.8)^{D/6 {\rm Mpc}}$ for a nucleon traversing a distance $D$. 
Since no active galactic nucleus-like sources are known to exist within 100
Mpc of the earth, the energy requirement for a proton arriving at the
earth with a
super-GZK energy is unrealistically high. 
Nevertheless, several experiments have reported handfuls of events above
$10^{20}$~eV (see e.g. \cite{crrev}). 
While data from HiRes brought these results into question,
a recent reevaluation of the AGASA data seems to confirm a violation of
the GZK cutoff. The issue will be solved soon conclusively by the
Pierre Auger observatory.
Among the solutions proposed for the origin of EECR's,
a rather conservative and economical scenario involves cosmic ray 
neutrinos which scatter resonantly at the cosmic neutrino background (CNB) 
predicted by Standard Cosmology and produce Z-bosons \cite{Zburst}. 
These Z-bosons in turn decay to produce a highly boosted ``Z-burst'',
containing on average twenty photons and two nucleons above $\egzk$.
The photons and nucleons from Z-bursts produced within a distance of
50 to 100 Mpc
can reach the earth with enough energy to initiate the 
air-showers observed at $\sim 10^{20}$~eV.

The energy of the neutrino annihilating at the peak of the Z-pole is
\be{}
E_{\nu_j}^R=\frac{M_Z^2}{2 m_j}=4\,(\frac{{\rm eV}}{m_j})\,{\rm ZeV}.
\ee
Even allowing for energy fluctuations about mean values, 
it is clear that in the Z-burst model the relevant
neutrino mass cannot exceed $\sim 1$~eV.
On the other hand, the neutrino mass cannot be too light. Otherwise
the predicted primary energies will exceed the observed
event energies and the primary neutrino flux will be pushed
to unattractively higher energies.
In this way,
one obtains a rough lower limit on the
neutrino mass of $\sim 0.1$~eV for the Z-burst model 
(with allowance made for an order of magnitude energy-loss 
for those secondaries traversing 50 to 100 Mpc). 
A detailed
comparison of the predicted proton spectrum with the observed EECR spectrum
in \cite{ringwald} yields a value of 
$m_{\nu}=0.26^{+0.20}_{-0.14}$~eV for extragalactic 
halo origin of the power-like part of the spectrum.

A necessary condition for the viability of this 
model is a sufficient flux of neutrinos at $\gsim 10^{21}$ eV.
Since this condition seems challenging, any increase of the Z-burst rate 
is helpful, 
that ameliorates slightly the formidable flux requirement.
If the neutrinos are mass degenerate, then a further consequence is that
the Z-burst rate at $E_R$ is three times what it would be 
without degeneracy. 
If the neutrino is a Majorana particle,
a factor of two more is gained in the Z-burst rate relative 
to the Dirac neutrino case since the two active helicity states 
of the relativistic CNB
depolarize upon cooling to populate all spin states.

Moreover the viability of the Z-burst model is enhanced if the CNB neutrinos 
cluster in our matter-rich vicinity of the universe.
For smaller scales, the Pauli blocking of identical
neutrinos sets a limit on density enhancement.
With a  virial velocity within our Galactic halo 
of a couple hundred km/s,
it appears that Pauli blocking allows significant clustering on the
scale of our Galactic halo only if $m_j \gsim 0.5$~eV.
For rich clusters of galaxies, the virial velocities are a thousand km/s
or more. Thus significant clustering
on scales of tens of Mpc is not excluded
for $m_j \gsim 0.3$~eV. An interesting possibility 
is, that our nearest Super Cluster, Virgo, contains a large neutrino 
overdensity. In such a case the EECRs we observe are products of Z-bursts
occuring in Virgo, which are focussed by our Galactic wind onto earth, 
producing at the same time an apparently isotropic sky-map for the observed 
events \cite{ma}.

Thus, if the Z-burst model turns out to be the correct 
explanation of EECRs, then
it is probable that neutrinos possess masses in the range 
$m_{\nu}\sim (0.1-1)$~eV. 
Mass-degenerate neutrino models are then likely. 
Consequences are
a value of $m_{ee}>0.01$~eV, and thus
a signal of $0\nu\beta\beta$ in next generation experiments
(assuming the neutrinos are Majorana particles), 
good prospects for a signal in the KATRIN experiment, 
and
a neutrino mass sufficiently large to affect the 
comological power spectrum, see fig. \ref{uhe}. 

\section*{Acknowledgements}
HP would like to thank the organizers of SUSY'02 for the kind invitation to 
this inspiring meeting. 
This work was supported by the
DOE grant no.\ DE-FG05-85ER40226 and the Bundesministerium 
f\"ur Bildung und Forschung (BMBF, Bonn, Germany) under the 
contract number 05HT1WWA2.

\end{document}